\documentclass[
  aps,          
  pra,          
  reprint,      
  amsmath,
  amssymb,
  floatfix,     
  longbibliography
]{revtex4-2}

\usepackage{graphicx}
\usepackage{bm}
\usepackage[colorlinks=true,linkcolor=blue,citecolor=blue,urlcolor=blue]{hyperref}

\begin{document}

\title{Counterfactual Quantum Sensing: \\ What Interaction-Free Measurement Can and Cannot Buy}

\author{Christoph F. Wildfeuer}
\email{christoph.wildfeuer@fhnw.ch}
\affiliation{FHNW University of Applied Sciences and Arts Northwestern Switzerland,
School of Engineering and Environment, Klosterzelgstrasse 2, CH-5210 Windisch, Switzerland}

\date{\today}

\begin{abstract}
Interaction-free measurement (IFM) infers the presence of an absorbing object from a photon that, in the counterfactual sense, never interacted with it, and is widely described as a route to minimally invasive sensing. We ask what it actually buys, in estimation-theoretic terms. Writing the Elitzur-Vaidman interferometer as a channel-estimation problem, its Fisher information for the transmissivity $T$ of an object in one arm is $[2T(1-T)]^{-1}$, exactly half that of direct transmission probing, and the two schemes deliver identical Fisher information per absorbed photon, $[T(1-T)^2]^{-1}$: for measuring how transparent something is, the interferometer buys nothing. The advantage lies in discrimination, and we show that what it requires is not that the object be opaque but that the competing hypothesis be the object's absence. Against empty space the Chernoff information per absorbed photon grows as $(8/\pi^2)\left[(1-\sqrt{T})/(1+\sqrt{T})\right]N\ln N$ in the number of Zeno cycles $N$, even for a weakly absorbing object, while between two partial transparencies it does not grow at all. Parasitic per-cycle loss $\epsilon$ caps the advantage: the conclusive interrogations per absorbed photon peak at $(0.2625/\epsilon)(1-\sqrt{T})/(1+\sqrt{T})$, attained at $N_{\mathrm{opt}}=1.5936/\epsilon$ for an opaque object, which identifies the loss per cycle as the figure of merit governing how far interaction-free sensing can be pushed. Finally, the negative result is not special to the interferometer: for a single photon meeting a memoryless, non-dispersive object any number of times through arbitrary fixed optics, the accessible quantum Fisher information never exceeds that of the same incident flux spent on independent single-pass probes, and is generically far below it. This recovers the bound of Massar, Mitchison and Pironio for this class, by a short argument that also identifies when it is tight.
\end{abstract}

\maketitle

\section{Introduction}

Interaction-free measurement is the observation that an absorbing object placed in one arm of an interferometer alters the output statistics even in runs in which it absorbs nothing. Elitzur and Vaidman made the point vivid with a bomb that any single photon would detonate, and showed that its presence can nonetheless be certified with finite probability by a photon that was never absorbed \cite{E-V}. Kwiat \emph{et al.} realised the effect in the laboratory \cite{Kwiat1995}, then pushed its efficiency towards unity \cite{Kwiat1999}. Their method was to interrogate the object through a chain of weakly rotating cycles: each pass presents only a small amplitude to the object, and the quantum Zeno effect suppresses the absorption. Whether such a measurement is properly called counterfactual has been debated at length \cite{Vaidman2003,Vaidman2019}; nothing below depends on the answer.

The practical appeal is that a fragile sample might be interrogated without being damaged, and the idea has been carried into imaging \cite{White1998,Yang2023}, spectroscopy \cite{Chen2021} and x-ray interrogation \cite{Cohen2024}. It is also often called \emph{supersensitive}. In quantum optical metrology that word has a settled technical meaning, namely phase sensitivity below the shot-noise limit, and is kept distinct from superresolution \cite{Dowling2008}. In the interaction-free literature it does two jobs at once. It can mean that the scheme extracts more information per probe, or that it deposits less energy in the sample. The two claims are independent, and only one of them survives. Careful versions say as much: Miller \emph{et al.} obtain sub-shot-noise \emph{phase} estimation in an induced-coherence interferometer, and state plainly that the gain comes from squeezing and not from the geometry of the interferometer \cite{Miller2021}.

Much of this ground was mapped early. Massar, Mitchison and Pironio derived bounds relating the number of absorbed photons to the sensitivity of an absorption measurement. Their class of strategies is a wide one, admitting arbitrary photon number, ancillas and interaction steps. They concluded that ``using the quantum mechanical properties of light does not offer significant advantages over more traditional schemes such as counting absorbed photons or simple interferometric procedures that use one photon at a time'' \cite{Massar2001}. Mitchison and Massar showed separately that two \emph{partial} transparencies cannot be discriminated absorption-free, while complete transparency can be told from any partial one \cite{Mitchison2001}; with Pironio they went on to bound the photon cost of the harder case \cite{Mitchison2002}. Facchi \emph{et al.} compared standard and quantum Zeno absorption tomography and found that, expressed per absorbed particle, the two Cram\'er-Rao bounds ``reduce to the same bound'' \cite{Facchi2002}.

Closest to the present work, Thomas \emph{et al.} studied semitransparent objects in interaction-free measurement by counting probe particles \cite{Thomas2014}. They reached both of the conclusions reached here, by other means: that for determining an unknown transparency such measurements ``lose as many probe particles as classical measurements or more'', and that ``interaction-free measurements with zero loss are possible if one of the samples is perfectly transparent''.

What this paper adds is quantitative. Section~\ref{sec:est} treats the Elitzur-Vaidman interferometer as an estimation problem: it is a factor of two \emph{worse} than direct probing, and exactly as good once we count the damage done. Section~\ref{sec:zeno} locates the advantage that survives, in discrimination rather than estimation, and gives in closed form the rate at which it grows and the condition it requires. Section~\ref{sec:loss} gives the limit that parasitic loss places on that advantage, also in closed form; this is the result an experiment can be designed against. Section~\ref{sec:thm} then shows that the negative result of Sec.~\ref{sec:est} is no accident of one interferometer but holds for every fixed multi-pass arrangement of a single photon, which recovers the bound of Ref.~\cite{Massar2001} for this class by a short argument and identifies when it is tight.

\section{Interaction-free measurement as a parameter-estimation problem}
\label{sec:est}

\begin{figure*}[t]
    \centering
    \includegraphics[width=\textwidth]{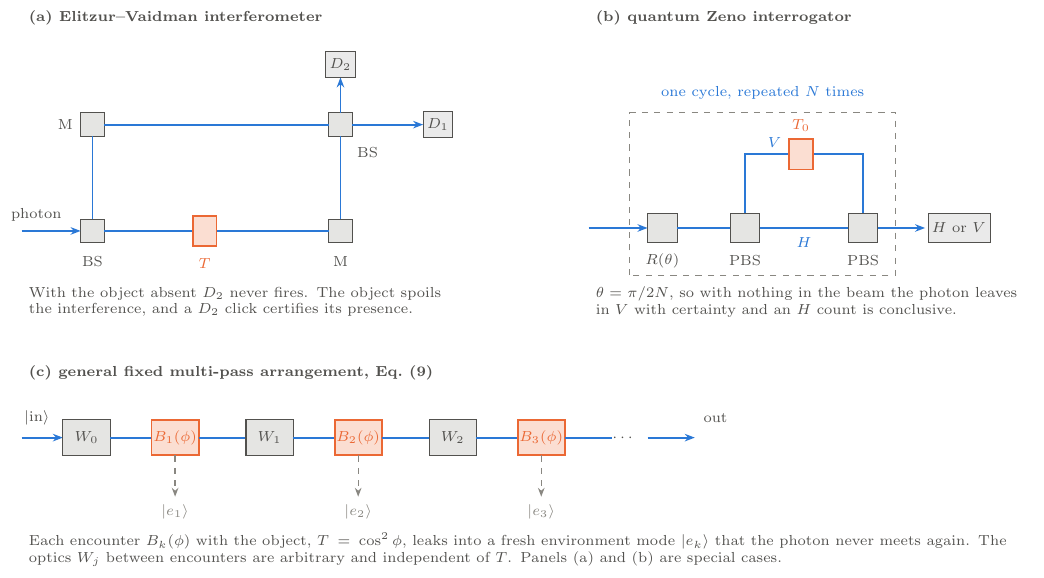}
    \caption{The three arrangements considered. (a) is the interferometer of Sec.~\ref{sec:est}, (b) the Zeno interrogator of Secs.~\ref{sec:zeno} and \ref{sec:loss}, and (c) the general class covered by the bound of Sec.~\ref{sec:thm}, of which the first two are special cases.}
    \label{fig:setup}
\end{figure*}

Let the object have intensity transmissivity $T\in[0,1]$ and amplitude transmission $t=\sqrt{T}$, with $T=1$ corresponding to no object at all. Writing each beam splitter as $\frac{1}{\sqrt{2}}\left(\begin{smallmatrix}1 & i\\ i & 1\end{smallmatrix}\right)$, the outcome probabilities of the Mach--Zehnder interferometer of Fig.~\ref{fig:setup}(a) are
\begin{align}
p_1(T) &= \tfrac{1}{4}\left(1+\sqrt{T}\right)^2, \label{eq:p1}\\
p_2(T) &= \tfrac{1}{4}\left(1-\sqrt{T}\right)^2, \label{eq:p2}\\
p_{\mathrm{abs}}(T) &= \tfrac{1}{2}\left(1-T\right), \label{eq:pabs}
\end{align}
where $p_1$ and $p_2$ refer to the bright and dark ports and $p_{\mathrm{abs}}$ is the probability that the object absorbs the photon. For $T=0$ these reduce to $\left(\tfrac14,\tfrac14,\tfrac12\right)$, recovering the Elitzur--Vaidman case, and for $T=1$ to $(1,0,0)$.

The relevant figure of merit is the Fisher information about $T$ carried by one interrogation, $F(T)=\sum_i p_i^{-1}\left(\partial_T p_i\right)^2$. Substituting Eqs.~(\ref{eq:p1})--(\ref{eq:pabs}), the $\left(1\pm\sqrt{T}\right)^2$ factors cancel and
\begin{equation}
F_{\mathrm{MZI}}(T) = \frac{1}{4T}+\frac{1}{4T}+\frac{1}{2(1-T)} = \frac{1}{2T(1-T)}.
\label{eq:Fmzi}
\end{equation}
Compare the obvious classical alternative: send the photon straight at the object and record whether it survives. That outcome is Bernoulli with success probability $T$, so
\begin{equation}
F_{\mathrm{dir}}(T) = \frac{1}{T(1-T)},
\label{eq:Fdir}
\end{equation}
which also saturates the quantum Fisher information for a single-photon probe of a pure-loss channel, since $T$ enters only through the populations of the $\{$transmitted, lost$\}$ basis. Hence
\begin{equation}
F_{\mathrm{MZI}}(T) = \tfrac{1}{2}\,F_{\mathrm{dir}}(T)
\label{eq:half}
\end{equation}
for every $T$: as an estimator of transmissivity the interferometer sits a factor of two below direct probing. The reason is simple, since only half the amplitude ever visits the object. It is worth saying out loud anyway, because it is the opposite of what ``supersensitivity'' suggests.

\begin{figure}[t]
    \centering
    \includegraphics[width=\columnwidth]{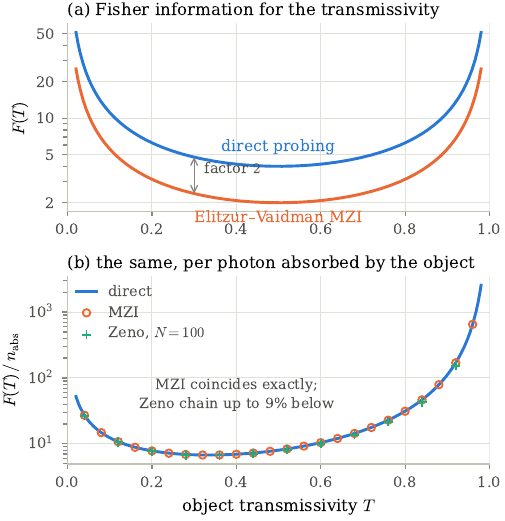}
    \caption{The central negative result. (a) Fisher information about the transmissivity $T$ of an object placed in one arm of the interferometer, Eq.~(\ref{eq:Fmzi}), against direct transmission probing, Eq.~(\ref{eq:Fdir}). The interferometer sits a factor of two below the direct scheme at every $T$. (b) The same quantities normalised by the number of photons the object absorbs. The Elitzur--Vaidman interferometer (circles) falls exactly on the direct-probing curve $[T(1-T)^2]^{-1}$ (solid). The $100$-cycle Zeno chain (crosses) lies between $1$ and $9\%$ below it. That shortfall is real and not a defect of the measurement: the classical Fisher information plotted here equals the quantum Fisher information of the reduced photon state, so the $\{H,V,\mathrm{lost}\}$ measurement is already optimal. Cycling discards information, as the bound of Sec.~\ref{sec:thm} allows it to.}
    \label{fig:fisher}
\end{figure}

The natural rebuttal is that the comparison is unfair: the point of IFM is that it absorbs fewer photons. That is true, and it is also exactly compensated. Throughout, $n_{\mathrm{abs}}$ denotes the mean number of photons the object absorbs in one interrogation, whatever the scheme. Here the object absorbs $n_{\mathrm{MZI}}=(1-T)/2$ photons per interrogation in the interferometer and $n_{\mathrm{dir}}=(1-T)$ in the direct scheme, so
\begin{equation}
\frac{F_{\mathrm{MZI}}}{n_{\mathrm{MZI}}} = \frac{F_{\mathrm{dir}}}{n_{\mathrm{dir}}} = \frac{1}{T(1-T)^2}.
\label{eq:invariance}
\end{equation}
The information gained per unit of damage done is identical, as Fig.~\ref{fig:fisher} shows. The interferometer buys its gentleness at precisely the exchange rate one would expect, and nothing more.

\section{The Zeno chain and where the advantage does live}
\label{sec:zeno}

\begin{figure}[t]
    \centering
    \includegraphics[width=\columnwidth]{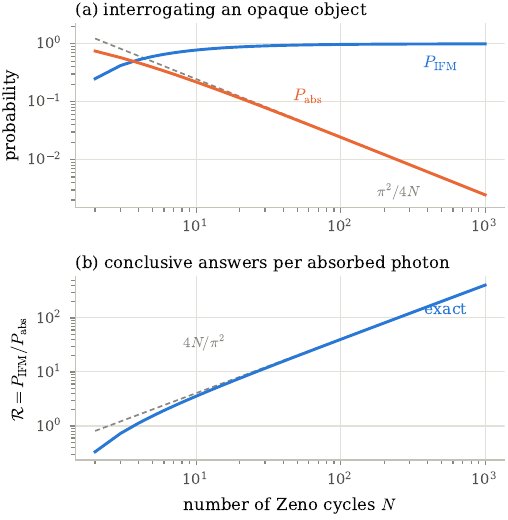}
    \caption{Quantum Zeno interrogation of an opaque object. (a) The probability $P_{\mathrm{IFM}}$ of a conclusive interaction-free answer, Eq.~(\ref{eq:zeno}), and the probability $P_{\mathrm{abs}}$ that the object absorbs the photon; the dashed line is the asymptote $\pi^2/4N$. (b) Their ratio $\mathcal{R}$, the number of conclusive interrogations per absorbed photon, approaching $4N/\pi^2$ (dashed) from below by the constant $\tfrac12$ of Eq.~(\ref{eq:ratio}). The growth is unbounded in a lossless chain.}
    \label{fig:zeno}
\end{figure}

The high-efficiency scheme of Kwiat \emph{et al.} \cite{Kwiat1999} cycles the photon $N$ times through a weakly rotating element, the object acting on one polarisation component each pass [Fig.~\ref{fig:setup}(b)]. With rotation angle $\theta=\pi/(2N)$, one cycle is
\begin{equation}
A = \begin{pmatrix} 1 & 0\\ 0 & \sqrt{T}\end{pmatrix}\begin{pmatrix}\cos\theta & -\sin\theta\\ \sin\theta & \cos\theta\end{pmatrix},
\label{eq:cycle}
\end{equation}
acting on the polarisation amplitudes, and the state after the chain is $A^N(1,0)^{\mathsf{T}}$. We write $P_H$ and $P_V$ for the probabilities that the photon leaves unrotated or rotated, and $P_{\mathrm{abs}}$ for the probability that the object absorbs it. With the object absent ($T=1$) the accumulated rotation is $N\theta=\pi/2$ and the photon emerges rotated with certainty. With an opaque object ($T=0$) the rotated amplitude is removed at every pass and the photon survives unrotated with probability
\begin{equation}
P_{\mathrm{IFM}}(N) = \cos^{2N}\!\left(\frac{\pi}{2N}\right) = 1-\frac{\pi^2}{4N}+\mathcal{O}\!\left(N^{-2}\right),
\label{eq:zeno}
\end{equation}
so the photons deposited in the object fall as $\pi^2/(4N)$ and the ratio of conclusive interrogations to absorbed photons grows without bound,
\begin{equation}
\mathcal{R}(N) \equiv \frac{P_{\mathrm{IFM}}}{P_{\mathrm{abs}}} = \frac{4N}{\pi^2}-\frac{1}{2}+\mathcal{O}\!\left(N^{-1}\right),
\label{eq:ratio}
\end{equation}
which we confirm numerically: $\mathcal{R}=40.03$ at $N=100$ and $\mathcal{R}=404.78$ at $N=1000$, against $4N/\pi^2-\tfrac12 = 40.03$ and $404.78$. Both limits are plotted in Fig.~\ref{fig:zeno}.

Equation~(\ref{eq:ratio}) is often read as the supersensitivity the literature promises, and it cannot be. The information-per-damage exchange rate for \emph{estimating} a transmissivity is fixed for the interferometer, Eq.~(\ref{eq:invariance}), and bounded for every multi-pass arrangement, as Sec.~\ref{sec:thm} shows. What Eq.~(\ref{eq:ratio}) describes is a different task, \emph{discriminating} between two hypotheses about the object, and it is only there that the chain wins. Anisimov \emph{et al.} recast Zeno interrogation in exactly these terms, as a statistical test between a clear path and a blocked one \cite{Anisimov2010}, and Zhou and Yung analysed it as a channel-discrimination problem at finite cycle number, finding that entangled inputs give no advantage over single photons \cite{Zhou2017}.

Equation~(\ref{eq:zeno}) was derived at $T=0$, so it is natural to credit the win to the object being opaque. That is wrong, and the error matters: it would restrict the advantage to far fewer samples than actually enjoy it. What the win really needs is that one of the two hypotheses be the object's \emph{absence}.

\begin{figure}[t]
    \centering
    \includegraphics[width=\columnwidth]{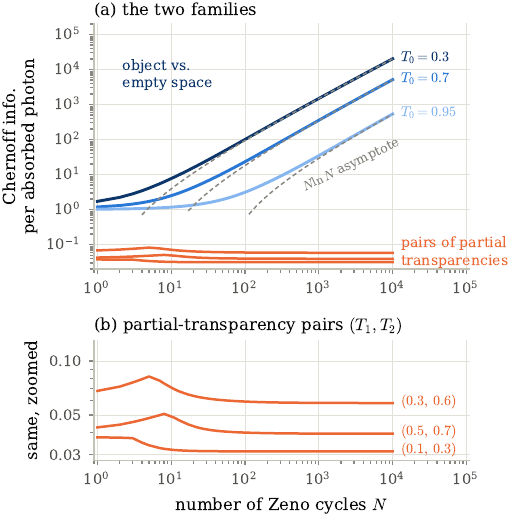}
    \caption{What the discrimination advantage requires. Chernoff information for an $N$-cycle chain, normalised by the photons the object absorbs under whichever of the two hypotheses absorbs more. (a) Blue: an object of transmissivity $T_0$ interrogated against empty space; dashed lines are the closed-form asymptote of Eq.~(\ref{eq:NlogN}), which the exact curves join once $N$ exceeds a few tens. The rate grows as $N\ln N$ for every $T_0$ shown, none of which is close to opaque. Orange: two partial transparencies interrogated against each other; the rate does not grow. The case $T_0=0$ is omitted from the blue family because the two outcome distributions then have disjoint support, a single interrogation decides, and the Chernoff information is infinite. (b) The orange family on its own scale, where the three pairs $(T_1,T_2)$ can be separated and their flatness in $N$ checked; note that the pair $(0.3,0.6)$ outperforms $(0.5,0.7)$, so the rate is set by the separation of the two candidates and not by how much either absorbs.}
    \label{fig:regimes}
\end{figure}

The reason is best put in old-fashioned terms: the chain is a \emph{null experiment}. With nothing in the beam the apparatus is tuned so that one detector never fires --- the $N$ rotations sum to exactly $\pi/2$ and the photon leaves in $|V\rangle$ with certainty, so $P(H\,|\,\mathrm{absent})=0$ \emph{exactly}, not merely small. A count in the $H$ port is then not evidence but proof. What the Zeno effect contributes is that this null can be made deep while the sample is barely touched: the absorbed fraction falls as $\pi^2/4N$ while the null stays exact, so a conclusive click costs almost no damage. Note where that zero comes from. It is a property of the empty apparatus, not of the object, which is why the advantage survives at $T_0=0.95$ in Fig.~\ref{fig:regimes}. Any object at all spoils the perfect rotation and leaks amplitude into the forbidden port. It does not have to be opaque; it only has to be there.

Estimation has no null. Asking how transparent an object is means distinguishing $T$ from $T+\mathrm{d}T$, and every outcome then carries non-zero probability under both hypotheses. Nothing is forbidden, so nothing is ever proof, and each run contributes a statistical increment rather than a certainty --- which is to say one is back to counting photons, the currency that Eq.~(\ref{eq:invariance}) and the bound of Sec.~\ref{sec:thm} govern. Two partial transparencies are in the same position: with $T_1$ and $T_2$ both below unity, both leak into both ports, and that is why the family in Fig.~\ref{fig:regimes}(b) is flat. In information-theoretic terms an outcome of probability zero under one hypothesis makes the divergence between the two distributions unbounded, so a single trial can carry unbounded information; once both distributions have full support every divergence is finite and the information per trial is capped.

Figure~\ref{fig:regimes} makes the separation quantitative. We compare two hypotheses by their Chernoff information $C$, the exponent that governs the error probability of an optimal test, and normalise it as before by the photons the object absorbs --- specifically by the absorption under whichever hypothesis absorbs more, since that is the damage the experiment must budget for. Against empty space, an object of transmissivity $T_0$ gives a rate that grows without bound for every $T_0<1$. This includes $T_0=0.95$, which absorbs five per cent of what reaches it and is in no sense opaque. Against another partial transparency $T_1$ versus $T_2$, both below unity, the rate does not grow at all, even when one of the two candidates is perfectly opaque. What the advantage tracks is conclusiveness, not opacity.

The growth law can be given in closed form, and it is worth doing so because a fitted exponent is misleading here. With nothing in the beam the photon leaves in $|V\rangle$ with certainty, so $V$ is the only outcome with support under both hypotheses, the Chernoff exponent collapses to its boundary value, and $C=-\ln P_V$ \emph{exactly}. The chain settles to a steady $V$ amplitude $\sqrt{T}\,\theta/(1-\sqrt{T})$ before each encounter, giving
\begin{align}
P_V &\longrightarrow \frac{T\pi^2}{4N^2\left(1-\sqrt{T}\right)^2}, \label{eq:PVasym}\\
n_{\mathrm{abs}} &\longrightarrow \frac{\pi^2}{4N}\,\frac{1+\sqrt{T}}{1-\sqrt{T}}, \label{eq:nabsasym}
\end{align}
so that $C=2\ln N+\ln\!\left[4(1-\sqrt{T})^2/T\pi^2\right]$ and
\begin{equation}
\frac{C}{n_{\mathrm{abs}}} \;\longrightarrow\; \frac{8}{\pi^2}\,\frac{1-\sqrt{T}}{1+\sqrt{T}}\;N\ln N .
\label{eq:NlogN}
\end{equation}
The rate grows as $N\ln N$, marginally faster than linearly. Both prefactors check out: at $N=10^8$ we find $N n_{\mathrm{abs}} = 8.447,\,27.74,\,192.42$ for $T=0.3,\,0.7,\,0.95$ against the $8.444,\,27.74,\,192.43$ of Eq.~(\ref{eq:nabsasym}), and the additive constant in $C$ agrees to three decimals or better. We stress the closed forms rather than a fit because the subleading constant in $C$ is large and negative, so a log-log slope measured over any accessible range of $N$ reads about $1.1$ rather than the asymptotic $1$; the exponent is better read off Eqs.~(\ref{eq:PVasym})--(\ref{eq:nabsasym}) than from the curves.

The null is exact only in the model, and it is worth saying what survives when it is not. Real polarising optics have a finite extinction ratio; Kwiat \emph{et al.} report crosstalk of order one per cent \cite{Kwiat1999}. Let a fraction $\delta$ of the amplitude leak into the wrong port, so that $P(H\,|\,\mathrm{absent})=\delta$ rather than zero. Two things change, and only one of them matters. An $H$ count is no longer proof: it carries a likelihood ratio $P(H\,|\,\mathrm{present})/\delta$, close to $1/\delta$ for a good chain --- about $10^2$ at one per cent crosstalk and $10^3$ at one part in $10^3$. Those are odds of a hundred to one from a single click, which is not certainty but is not far from it. Second, the Chernoff information no longer grows without bound. For this model of the imperfection we find numerically that it saturates at
\begin{equation}
C_\infty = -\ln\!\left[2\sqrt{\delta(1-\delta)}\,\right] \;\xrightarrow{\;\delta\ll1\;}\; \ln\frac{1}{2\sqrt{\delta}} ,
\label{eq:Cinf}
\end{equation}
which the exact chain approaches from above, agreeing to five significant figures at $N=10^6$ for $\delta=10^{-2},10^{-3}$ and $10^{-4}$ and for $T_0=0,\,0.3$ and $0.7$ alike.

The constant in Eq.~(\ref{eq:Cinf}) is specific to output-port extinction, in which a fraction $\delta$ is exchanged \emph{between} the ports, and should not be read as universal. Imperfections that leave $P(V\,|\,\mathrm{present})\to0$ --- a per-cycle retardance error, or a dark count in the $H$ channel --- saturate instead at $\ln(1/\delta)$, three times larger at $\delta=10^{-2}$. Finite detection efficiency is a separate and often tighter ceiling, and one the null argument depends on more than it depends on crosstalk: $C=-\ln P_V$ requires the no-click channel as well as the $H$ channel to be forbidden when the object is absent, which needs unit efficiency. At efficiency $\eta$ the exponent saturates at $\ln[1/(1-\eta)]$, which is $4.61$ at $\eta=0.99$ and $2.30$ at $\eta=0.9$.

What all four models share is the important part. The rate per absorbed photon continues to grow \emph{linearly} in $N$; only the logarithm freezes, $\ln N$ being replaced by a constant set by whichever imperfection dominates. The damage advantage, which is the useful part, survives imperfect optics and imperfect detectors alike. The logarithmic bonus does not, and at one per cent crosstalk it is gone by $N$ of order ten for $T_0=0.3$, though only near $N\simeq300$ for $T_0=0.95$, where the ideal chain has its own transient of $(1-\sqrt{T_0})^{-1}\simeq40$ cycles to get through first.

Neither half of this is new as a statement about particles. Mitchison and Massar proved the corresponding result for absorption-free discrimination --- complete transparency can be told from any partial transparency, two partial transparencies cannot \cite{Mitchison2001} --- and Thomas \emph{et al.} reached the same conclusion by counting probe particles through a semitransparent sample \cite{Thomas2014}. What Fig.~\ref{fig:regimes} adds is the rate at which the two cases separate, in a quantity that bounds every measurement rather than a particular one. The flat family in that figure is the case whose photon cost Mitchison, Massar and Pironio bounded \cite{Mitchison2002}. This also explains why estimation is the harder problem. Estimating a continuous $T$ means discriminating between neighbouring partial transparencies, and that is precisely the case with no conclusive outcome, where the bound of Sec.~\ref{sec:thm} binds.

The discrimination task is naturally posed in Bayesian terms, and the single-cycle case is instructive. Let $B$ be the event that the object is present, $\overline{B}$ its absence, and $D_i$ a click in port $i$. Suppose an opaque object is present with prior probability $P(B)=\tfrac12$. From Eqs.~(\ref{eq:p1})--(\ref{eq:pabs}) at $T=0$, $P(D_i|B)=\tfrac14$ for each port, while $P(D_1|\overline{B})=1$ and $P(D_2|\overline{B})=0$. Hence $P(D_1)=\tfrac58$ and $P(D_2)=\tfrac18$, and
\begin{equation}
P(B|D_1)=\frac{1}{5},\qquad P(B|D_2)=1 .
\label{eq:bayes}
\end{equation}
A click in the dark port is conclusive. A click in the bright port is inconclusive, but it is not uninformative: it moves the posterior from $\tfrac12$ down to $\tfrac15$. Informal accounts often miss that second point, though the numbers of course depend on the prior assumed. In an $N$-cycle chain the conclusive outcome still carries posterior unity, while the absorbed fraction falls as $\pi^2/4N$. That is the content of Eq.~(\ref{eq:ratio}).

The supersensitivity claim can now be stated precisely, and it depends entirely on the question asked. Ask how transparent an object is, and interaction-free measurement gains nothing: the information per absorbed photon is the same for every scheme in the family. Ask whether the object is there at all, and it gains a factor that grows as $N\ln N$, however weakly the object absorbs. Ask which of two semitransparent objects sits in the beam, and it gains nothing again.

The practical rule follows. Use interaction-free measurement to protect a fragile sample while checking whether something is present. Do not use it to measure how much is present, or to tell one semitransparent sample from another.

\section{The loss-limited optimum}
\label{sec:loss}

Equation~(\ref{eq:ratio}) grows without bound only in a lossless chain. A lossy Zeno interrogator was studied by Anisimov \emph{et al.}, who modelled loss by a beam splitter in the interrogated arm and reported a non-monotonic dependence of the transmission probability on it \cite{Anisimov2010}. They did not extract an optimal cycle number, and did not compare conclusive detections with absorbed photons; both follow in closed form below. Let each cycle carry a survival probability $1-\epsilon$ from imperfect optics, independent of the object. A photon lost this way at cycle $j$ is also never absorbed by the object, so both channels are attenuated: the conclusive one over the whole chain, the absorbed one only up to the cycle where absorption happens. With $u=(1-\epsilon)\cos^2\theta$,
\begin{equation}
\mathcal{R}_\epsilon(N) = \frac{u^N\,(1-u)}{\sin^2\theta\;\bigl(1-u^N\bigr)}
\;\simeq\; \frac{4N}{\pi^2}\,\frac{x}{e^{x}-1},\qquad x=N\epsilon .
\label{eq:lossy}
\end{equation}
Maximising the large-$N$ form gives the transcendental condition $2\left(e^{x}-1\right)=x\,e^{x}$, whose root is $x^\star=1.59362$, so that
\begin{align}
N_{\mathrm{opt}} &= \frac{x^\star}{\epsilon} = \frac{1.5936}{\epsilon}, \nonumber\\
\mathcal{R}_{\max} &= \frac{4}{\pi^2}\,\frac{(x^\star)^2}{e^{x^\star}-1}\,\frac{1}{\epsilon} = \frac{0.26247}{\epsilon}.
\label{eq:opt}
\end{align}
Maximising the exact expression in Eq.~(\ref{eq:lossy}) over integer $N$ reproduces this over four decades: for $\epsilon = 10^{-2},10^{-3},10^{-4},10^{-5}$ we find $N_{\mathrm{opt}} = 160,\,1594,\,15937,\,159362$ and $\mathcal{R}_{\max} = 25.73,\,261.95,\,2624.1,\,26246$, against the predicted $159,\,1594,\,15936,\,159362$ and $26.25,\,262.47,\,2624.7,\,26247$. The extension to a partially transmitting object is asymptotic rather than exact, and the condition matters. Because the parasitic factor $\sqrt{1-\epsilon}\,\mathbb{1}$ is a scalar it commutes with everything, so
\begin{equation}
P_H^{\epsilon}(N) = (1-\epsilon)^N P_H^{(0)}(N), \quad
n_{\mathrm{abs}}^{\epsilon}(N) = \sum_{k=1}^{N}(1-\epsilon)^{k-1}d_k
\label{eq:factor}
\end{equation}
exactly, with $d_k$ the lossless per-cycle absorption. Replacing $\pi^2/4$ by the constant of Eq.~(\ref{eq:nabsasym}) then gives
\begin{equation}
N_{\mathrm{opt}} \simeq \frac{x^\star}{\epsilon}, \qquad
\mathcal{R}_{\max} \simeq \frac{0.26247}{\epsilon}\,\frac{1-\sqrt{T_0}}{1+\sqrt{T_0}} ,
\label{eq:optT}
\end{equation}
but only when $d_k$ has reached its steady value and $P_H^{(0)}\to1$, both of which require $N\gg(1-\sqrt{T_0})^{-1}$, that is $\epsilon\ll1-\sqrt{T_0}$. Outside that regime the optimum moves. Exhaustive maximisation of the exact chain over integer $N$ gives

\begin{center}
\begin{tabular}{lccccc}
\hline\hline
$\epsilon$ & $T_0{=}0$ & $0.3$ & $0.7$ & $0.95$ & $0.99$ \\
\hline
$10^{-2}$ & $160$ & $161$ & $165$ & $195$ & $230$ \\
$10^{-3}$ & $1594$ & $1595$ & $1599$ & $1636$ & $1802$ \\
\hline\hline
\end{tabular}
\end{center}

\noindent so that at $\epsilon=10^{-2}$ the peak has moved by $44\%$ by $T_0=0.99$. The departure is governed by $\mu=\epsilon/(1-\sqrt{T_0})$: $N_{\mathrm{opt}}$ exceeds $x^\star/\epsilon$ by $22\%$ at $\mu\simeq0.4$ and by $44\%$ at $\mu\simeq2$. Equation~(\ref{eq:optT}) is therefore the small-$\mu$ limit, and the leading term $N_{\mathrm{opt}}=x^\star/\epsilon$ is exact only for an opaque object, where the earlier table is reproduced to the unit digit.

\begin{figure}[t]
    \centering
    \includegraphics[width=\columnwidth]{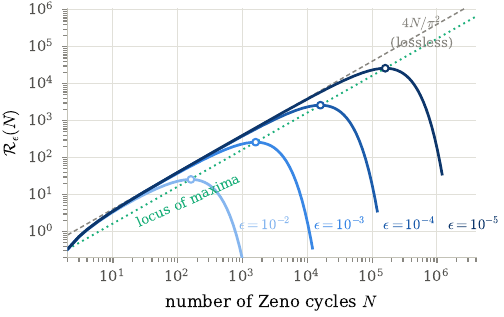}
    \caption{The loss-limited optimum. Conclusive interrogations per absorbed photon, $\mathcal{R}_\epsilon(N)$ of Eq.~(\ref{eq:lossy}), for four values of the parasitic per-cycle loss $\epsilon$. Open circles mark the maxima, located on the integer grid; the dotted line is their locus, reached at $N_{\mathrm{opt}}=x^\star/\epsilon$, and the dashed line is the lossless asymptote $4N/\pi^2$. Cycling past $N_{\mathrm{opt}}$ is actively harmful: the chain leaks the photon before it can answer.}
    \label{fig:loss}
\end{figure}

Figure~\ref{fig:loss} shows the four curves and the locus their maxima trace out for an opaque object. The practical message is plain: the loss per cycle sets the damage advantage of interaction-free sensing, and nothing else does. Improve $\epsilon$ by a factor of ten and you protect the sample ten times better. Note also that Eq.~(\ref{eq:optT}) gives $\mathcal{R}_{\max}<1$ --- fewer conclusive interrogations than absorbed photons, hence no advantage at all --- whenever $(1-\sqrt{T_0})/(1+\sqrt{T_0})<\epsilon/0.26247$; at $\epsilon=10^{-2}$ and $T_0=0.95$ the exact chain gives $\mathcal{R}_{\max}=0.32$. Multi-pass electron microscopy runs into the same trade-off, where the number of passes is limited by loss per pass rather than by the physics of the interrogation \cite{Kruit2016,Juffmann2016}.

\section{How far the negative result generalises}
\label{sec:thm}

Equation~(\ref{eq:invariance}) is a statement about two particular interferometers. We now show that no arrangement of a single photon does better, so that the exchange rate is a property of the problem and not of the device.

\emph{Setup.} Put $T=\cos^2\phi$ and dilate the object to a beam splitter of angle $\phi$ coupling the object mode $|o\rangle$ to a fresh vacuum mode $|e_k\rangle$ at the $k$th of $K$ encounters, so that the global photon-plus-environment state stays pure. An arbitrary scheme, sketched in Fig.~\ref{fig:setup}(c), is then
\begin{equation}
U(\phi) = W_K B_K(\phi)\, W_{K-1} \cdots W_1 B_1(\phi)\, W_0 ,
\label{eq:general}
\end{equation}
with $B_k(\phi)=e^{-i\phi J_k}$ the $k$th encounter, $J_k = |o\rangle\langle e_k| + |e_k\rangle\langle o|$ in the single-excitation subspace, and the $W_j$ arbitrary optics independent of $T$. Multi-pass geometries of this kind are standard in optical metrology; a Fabry--Perot resonator is the canonical example \cite{Wildfeuer2009}. Writing $|\psi_k\rangle$ for the global state after the $k$th encounter and the optics that follow it, with $|\psi_0\rangle=W_0|\mathrm{in}\rangle$, let $v_k=\langle o|\psi_{k-1}\rangle$ be the amplitude incident on the object at pass $k$, so that the object absorbs
\begin{equation}
n_{\mathrm{abs}} = (1-T)\sum_{k=1}^{K}|v_k|^2
\label{eq:nabs-general}
\end{equation}
photons on average. The argument needs three physical assumptions, which we state and use below.

\emph{Lemma.} $\bigl\|\partial_\phi|\psi\rangle\bigr\|^2=\sum_k |v_k|^2$. We induct on $K$. From $|\psi_K\rangle = W_K B_K|\psi_{K-1}\rangle$, the product rule, and the unitarity of $W_K$ and $B_K$ (using $[J_K,B_K]=0$),
\begin{align}
\bigl\|\partial_\phi\psi_K\bigr\|^2 ={}& \bigl\|J_K\psi_{K-1}\bigr\|^2 + \bigl\|\partial_\phi\psi_{K-1}\bigr\|^2 \nonumber\\
&{}- 2\,\mathrm{Im}\,\bigl\langle J_K\psi_{K-1}\big|\partial_\phi\psi_{K-1}\bigr\rangle .
\label{eq:induction}
\end{align}
Because $e_K$ is a mode the photon has not yet touched, $\langle e_K|\psi_{K-1}\rangle=0$, so $J_K|\psi_{K-1}\rangle = \langle e_K|\psi_{K-1}\rangle|o\rangle + v_K|e_K\rangle = v_K|e_K\rangle$ and the first term is $|v_K|^2$. That identity holds for all $\phi$, so it may be differentiated: $\partial_\phi|\psi_{K-1}\rangle$ likewise has no $e_K$ component and the cross term vanishes. This is a statement about the support of the state and is insensitive to the phases of the amplitudes. The second term is $\sum_{k<K}|v_k|^2$ by induction, and the base case is $\|\partial_\phi\psi_0\|^2=0$ because $|\psi_0\rangle = W_0|\mathrm{in}\rangle$ does not depend on $\phi$. $\square$

\emph{Purification identity.} The generator $G=iU^\dagger\partial_\phi U$ has vanishing expectation. Peeling the tail unitaries, which cancel between bra and ket,
\begin{equation}
\langle\psi|\partial_\phi\psi\rangle = -i\sum_{k=1}^{K}\langle\psi_{k-1}|J_k|\psi_{k-1}\rangle = 0,
\label{eq:Gzero}
\end{equation}
each term vanishing by $\langle e_k|\psi_{k-1}\rangle=0$ --- the same freshness property that powers the Lemma. Hence $F_Q^{\mathrm{pur}}(\phi)=4\|\partial_\phi\psi\|^2$, and with $(\mathrm{d}T/\mathrm{d}\phi)^2=\sin^2 2\phi=4T(1-T)$ the Lemma gives
\begin{equation}
F_Q^{\mathrm{pur}}(T) = \frac{\bar n_{\mathrm{tot}}}{T(1-T)},
\qquad \bar n_{\mathrm{tot}}=\sum_{k=1}^{K}|v_k|^2,
\label{eq:pur}
\end{equation}
exactly, for every $K$, every choice of intervening optics and every $T\in(0,1)$. $\square$

\emph{Theorem.} The photon-plus-environment state is pure, and the state an experimenter actually holds is its reduction, obtained by discarding the environment. Write $F_Q$ for the quantum Fisher information of that reduced photon state --- the quantity any apparatus is limited by. Since the quantum Fisher information is monotone under the partial trace, $F_Q\le F_Q^{\mathrm{pur}}$, and with $n_{\mathrm{abs}}=(1-T)\bar n_{\mathrm{tot}}$,
\begin{equation}
\boxed{\;\frac{F_Q(T)}{n_{\mathrm{abs}}} \;\le\; \frac{1}{T(1-T)^2}\;}
\label{eq:nogo}
\end{equation}
for every fixed multi-pass arrangement. $\square$

Equation~(\ref{eq:pur}) says that $\bar n_{\mathrm{tot}}$ is the right way to count the light an object sees, and Eq.~(\ref{eq:nogo}) then has a direct reading: \emph{a $K$-pass scheme is never more informative than the same incident flux spent on independent single-pass probes}, since a single pass saturates $\bar n/[T(1-T)]$. Equality is attainable --- trivially for $K=1$, and also for multi-pass arrangements whose optics route the photon away after its first encounter, which are single-pass in disguise; numerical optimisation of the intervening optics reaches the bound at $K=2$ and $K=3$. For arrangements in which the photon genuinely returns to the object the inequality is strict, and often badly so. Ten passes with no intervening optics at $T=0.5$ reach only $4.9\%$ of the bound; five passes at $T=0.05$ reach $1.4\times10^{-4}$ of it. Repeatedly interrogating an object does not merely fail to help, it discards most of the information the light carries.

A word is needed on multi-pass microscopy, which reports the opposite. Juffmann \emph{et al.} measure a $4.8\pm0.8$~dB variance reduction for transmission measurements at constant damage \cite{Juffmann2016} --- a factor of three, in the same currency as Eq.~(\ref{eq:nogo}). There is no conflict, and the reason is instructive. Their probe is classical light, and a coherent state is a poor probe of loss: its Fisher information per incident photon is $1/T$, against the $[T(1-T)]^{-1}$ of Eq.~(\ref{eq:nogo}), so a classical single-pass measurement begins a factor $(1-T)^{-1}$ below the bound --- a factor of one hundred at the $T\simeq0.99$ typical of microscopy. Multi-passing classical light climbs inside that gap. For $m$ passes with shot-noise-limited detection of the transmitted intensity, the Fisher information per unit dose is $m^2T^{m-2}(1-T)/(1-T^m)$, which at $T=0.99$ rises from $1\%$ of the bound at $m=1$ to $65\%$ at $m=159$, and never crosses it: maximising over $T\in(0,1)$ and $m\le500$ gives $0.99$ of the bound. A threefold gain over a classical baseline is therefore exactly what should be expected. It is a gain \emph{towards} the single-photon limit, not past it.

The contrast with phase estimation is worth drawing, because it is the reason multi-passing has a good reputation. Yu \emph{et al.} analyse multi-pass interferometry in this same per-dose currency and find that for \emph{phase} the sequential strategies do beat the parallel ones \cite{Yu2026}. The asymmetry is not accidental. A phase is imprinted coherently and accumulates over passes, so the signal grows as $m$ while the dose grows as $m$; each encounter with a lossy object, by contrast, is an independent Bernoulli trial, and independent trials do not compound.

Two features of the bound are worth noting. It is stated in terms of $n_{\mathrm{abs}}$, the mean number of photons the object absorbs, which is a property of the apparatus rather than of a chosen dilation, so Eq.~(\ref{eq:nogo}) does not depend on how the environment is modelled --- only $F_Q^{\mathrm{pur}}$ would. And $T$ is throughout the transmissivity of one elementary encounter; a device that presents several thicknesses per pass is described by re-defining the encounter, not by re-using Eq.~(\ref{eq:nogo}) with the same $T$.

Fisher-information analysis of lossy interferometers is a mature subject, but loss there is almost always a nuisance that degrades phase estimation \cite{Lee2009}; here loss is the parameter being estimated. The route taken here --- dilate, evaluate the quantum Fisher information of the purification, then invoke monotonicity under the partial trace --- is Nair's \cite{Nair2018}, and Eq.~(\ref{eq:pur}) is his $\tilde{K}_\phi = 4\,\mathrm{diag}(N_k)$ written for one object met $K$ times rather than $K$ objects met once. The underlying constant is the Monras--Paris limit \cite{Monras2007}, which Fock states saturate \cite{Adesso2009}. Nair's scope is the ancilla-assisted \emph{parallel} strategy, and he closes by noting that sequential strategies remain to be examined; Eqs.~(\ref{eq:pur})--(\ref{eq:nogo}) settle the fixed sequential case at finite $K$, where the per-pass amplitudes $v_k$ do the bookkeeping and the answer is that re-using a photon buys nothing. Protocols reported to beat the limit, such as the time-reversed scheme of Wang \emph{et al.} \cite{Wang2024}, in fact reach it and stop. Massar, Mitchison and Pironio obtained an inequality of this kind for a broader class by other means \cite{Massar2001}. General arguments that adaptive strategies do not asymptotically outperform parallel ones \cite{DemkowiczDobrzanski2014,Pirandola2017} point the same way, but they count channel uses rather than incident energy and are proved for phase-type parameters.

Three points about scope.

First, the gap between $F_Q$ and $F_Q^{\mathrm{pur}}$ is the information locked in the environment: which pass absorbed the photon. That record is not available --- an absorbed photon is simply a missing count --- and for the Zeno chain the loss is complete in a strong sense. We find the classical Fisher information of the $\{H,V,\mathrm{lost}\}$ measurement to equal the quantum Fisher information of the reduced state at every $N$ and $T$ tested, so that measurement is optimal and the shortfall in Fig.~\ref{fig:fisher}(b) cannot be recovered by measuring better.

Second, the bound is on the quantum Fisher information, so no choice of measurement escapes it. Facchi \emph{et al.} reached a comparable conclusion for standard and Zeno tomography, but only for binomial statistics \cite{Facchi2002}. Equation~(\ref{eq:nogo}) drops that restriction and covers any fixed multi-pass arrangement.

Third, the three assumptions do real work. The optics $W_j$ must leave the environment modes $e_k$ alone. This is physically obvious, since one cannot redirect light the object has already absorbed, but it is what makes the induction close. The object must be memoryless, so that each pass meets fresh vacuum; an absorber interrogated faster than its own coherence time can re-emit, and the identity fails. And the object must be non-dispersive: a Kramers--Kronig phase on $\sqrt{T}$ builds up over passes and carries the problem outside the theorem. Anyone hoping to beat the exchange rate must break one of these three.

What Eq.~(\ref{eq:nogo}) does not cover is multi-photon probes in sequential arrangements, and adaptive schemes in which the optics $W_j$ are chosen conditionally on intermediate measurement outcomes. Both lie inside the broader class treated by Massar, Mitchison and Pironio \cite{Massar2001}; whether the short argument given here extends to them we have not determined. Generic asymptotic arguments that adaptive strategies do not outperform parallel ones \cite{DemkowiczDobrzanski2014,Pirandola2017} point the same way but count channel uses rather than incident energy, and so do not by themselves imply Eq.~(\ref{eq:nogo}).

\section{Conclusion}

Put in estimation-theoretic terms, the supersensitivity claim splits cleanly in two.

Measuring how transparent an object is, the Elitzur--Vaidman interferometer carries half the Fisher information of direct probing, Eq.~(\ref{eq:half}), and exactly the same Fisher information per absorbed photon, Eq.~(\ref{eq:invariance}). That exchange rate also bounds every fixed multi-pass arrangement of a single photon, Eq.~(\ref{eq:nogo}): no such scheme is more informative than the same incident flux spent on independent single-pass probes, and one that genuinely re-visits the object is strictly worse. On this measure there is no interaction-free advantage, and no rearrangement of the interferometer will produce one.

Deciding whether an object is present at all is a different matter. There the discrimination rate per absorbed photon grows as $N\ln N$ in the cycle number, Eq.~(\ref{eq:NlogN}), and the advantage is real, quantifiable, and limited only by the loss per cycle through Eq.~(\ref{eq:opt}). It does not require an opaque object, only that the competing hypothesis be empty space. That widens the class of samples it serves, and at the same time sharpens the boundary: between two partial transparencies the rate does not grow at all.

So interaction-free sensing is the right tool wherever the sample, and not the photon budget, is the scarce resource: fragile biological specimens \cite{Taylor2016}, photosensitive materials, x-ray interrogation \cite{Cohen2024}. It is the wrong tool wherever the goal is a more precise number. Massar, Mitchison and Pironio reached that conclusion two decades ago in a broader setting \cite{Massar2001}; Eqs.~(\ref{eq:pur})--(\ref{eq:nogo}) recover it for single-photon multi-pass schemes by a short argument, and say when it is tight.

Two questions remain open. We do not know whether the identity extends to multi-photon sequential and adaptive strategies; the inequality is known there \cite{Massar2001}, the equality is not. And Facchi \emph{et al.} report a real per-absorbed-particle advantage for Zeno tomography when the grey levels across a sample are unevenly distributed \cite{Facchi2002}. That is a claim about a prior over many pixels, not about estimating one parameter at fixed $T$, so the two results are probably answers to different questions --- but showing that would sharpen the boundary further. A word finally on testing Eq.~(\ref{eq:opt}), because we are not able to propose a clean protocol and would rather say so. Measuring $\mathcal{R}_{\max}$ directly requires separating the photons the object absorbs from those the optics lose, and at $\epsilon=10^{-2}$ the former are only a few per cent of all missing counts. The natural evasion is to test the \emph{position} of the maximum rather than its height, since an argmax is invariant under any $N$-independent multiplicative calibration error. That evasion fails on closer inspection. The obvious calibration-free estimator, $H_{\mathrm{present}}/(\mathrm{total}_{\mathrm{absent}}-\mathrm{total}_{\mathrm{present}})$, has its $(1-\epsilon)^N$ factors cancel by Eq.~(\ref{eq:factor}) and returns the \emph{lossless} ratio, which is monotone in $N$ and has no maximum at all: at $\epsilon=10^{-2}$ it runs $15.7,\,31.9,\,64.3,\,129,\,259,\,518$ over $N=40$ to $1280$ while the true $\mathcal{R}_\epsilon$ peaks at $25.7$ near $N=160$. Recovering $n_{\mathrm{abs}}^{\epsilon}$ from the counts requires the model one is trying to test. Nor is the argmax as robust as the invariance argument suggests: additive $N$-dependent backgrounds are not removed by it, and a per-cycle retardance error is unitary, hence invisible to a calibration based on the total count rate, while placing a term growing as $N^2$ in the $H$ port. We therefore leave the identification of a calibration-free observable that exhibits the predicted maximum as an open problem, and note that it is the main obstacle between Eq.~(\ref{eq:opt}) and an experiment.

\section*{Disclosure of AI use}

This work was developed with the assistance of Claude (Anthropic, model \texttt{claude-opus-5}), used in July 2026 for literature search and reference retrieval, algebraic and numerical derivation of the results in Secs.~\ref{sec:est}--\ref{sec:loss}, the induction proof of the Lemma in Sec.~\ref{sec:thm}, generation of the figures. All analytic results were checked by independent numerical computation, and all references were verified against publisher records. The author has verified the derivations reported here and is solely responsible for the content.

\begin{acknowledgments}
The author thanks Norbert Sauer for valuable discussions.
\end{acknowledgments}

\end{document}